\documentclass[12pt]{article}
\usepackage{amssymb,amsmath}
\usepackage{doublespace}
\usepackage{epsf}

\begin{document}
\begin{spacing}{1.0}

\title{THE KRAMERS PROBLEM IN THE ENERGY-DIFFUSION LIMITED REGIME}

\author{Jos\'{e} M. Sancho\\
Institute for Nonlinear Science 0402\\
University of California San Diego\\
La Jolla, California 92093-0402\\
and\\
Departament d'Estructura i Constituents de la Mat\`{e}ria\\
Universitat de Barcelona\\
Av. Diagonal 647, E-08028 Barcelona, Spain\\
\and
Aldo H. Romero\\
Department of Chemistry and Biochemistry 0340\\
and Department of Physics\\
University of California, San Diego\\
La Jolla, California 92093-0340\\
\and
Katja Lindenberg\\
Department of Chemistry and Biochemistry 0340\\
and Institute for Nonlinear Science\\
University of California San Diego\\
La Jolla, California 92093-0340}
 
\date{\today}
\maketitle
\end{spacing}
\begin{spacing}{1.5}

\begin{abstract}
The Kramers problem in the energy-diffusion limited regime of very low
friction is difficult to deal with analytically because of the
repeated recrossings of the barrier that typically occur before
an asymptotic rate constant is achieved.  Thus, the transmission
coefficient of particles over the potential barrier 
undergoes oscillatory behavior in time before settling
into a steady state.  Recently, Kohen and Tannor
[D. Kohen and D. J. Tannor, J. Chem. Phys. {\bf 103}, 6013
(1995)] developed a method based on the phase space distribution
function to calculate the transmission coefficient as a function
of time in the {\em high-friction} regime.  Here we formulate a parallel
method for the {\em low-friction} regime.  We find analytic results for the
full time and temperature dependence of the rate coefficient in this regime.
Our low-friction result at long times reproduces the equilibrium result
of Kramers at very low friction and extends it to higher friction
and lower temperatures below the turn-over region. Our results indicate
that the single most important quantity in determining the entire time
evolution of the transmission coefficient is the rate of energy loss
of a particle that starts above the barrier.   We test our results,
as well as those of Kohen and Tannor for the Kramers problem, against
detailed numerical simulations.  

\end{abstract}


\section{Introduction}
\label{intro}

The classic work of Kramers in 1940 \cite{Kramers} on reaction rates
in which the effect
of the solvent was taken into account in the form of a Markovian
dissipation
and Gaussian delta-correlated fluctuations has spawned an enormous and
important literature that continues to flourish \cite{Hanggi,Kohen}. 
Recent advances in
experimental methods that make it possible to monitor the progress of
chemical reactions on microscopic (even single-molecule single-event)
spatial and temporal scales \cite{Zewail,Castleman}
have led to a revival of interest in 
theoretical approaches that examine the problem on these detailed scales.

The enormous literature on the theoretical front since the appearance
of Kramers' seminal paper has evolved in many directions
that include more formal derivations of Kramers' own results,
extensions to larger parameter regimes and to non-Markovian
dissipation models,
generalizations to more complex potentials and to many degrees of freedom,
analysis of quantum effects, and application to specific experimental
systems. 

One such recent direction, developed by Kohen and Tannor
(KT) \cite{Kohen}, deals with 
the derivation of the rate coefficient in the Kramers problem and in the
more general Grote-Hynes problem \cite{Grote}
(that is, the Kramers problem extended to a non-Markovian
dissipative memory kernel) so as to obtain not only the asymptotic rate
constant but the behavior of the rate coefficient at all times.  This
derivation is
based on the reactive flux method \cite{Hanggi,Strauba,Straubb}
and allows the
interpretation of the
salient features of the time dependent rate coefficient $k(t)$ in terms of
the time evolution of the representative distribution functions that
originate at the top of the barrier, thus paralleling closely and usefully
the methods used in numerical simulations of the problem.  KT
analyze in detail the dependence on the dissipative memory kernel
of the time it takes the rate coefficient to reach its stationary
(equilibrium) value. Their extensive analysis,
however, does not cover a number of parameter regimes, nor do they check
their time-dependent results against numerical simulations.  We 
undertake such studies to complement their work.
Thus, we extend the parameter regime of analysis and
check some of their results as well as our new ones against
numerical simulations.

\begin{figure}[htb]
\begin{center}
\leavevmode
\epsfxsize = 3.2in
\epsffile{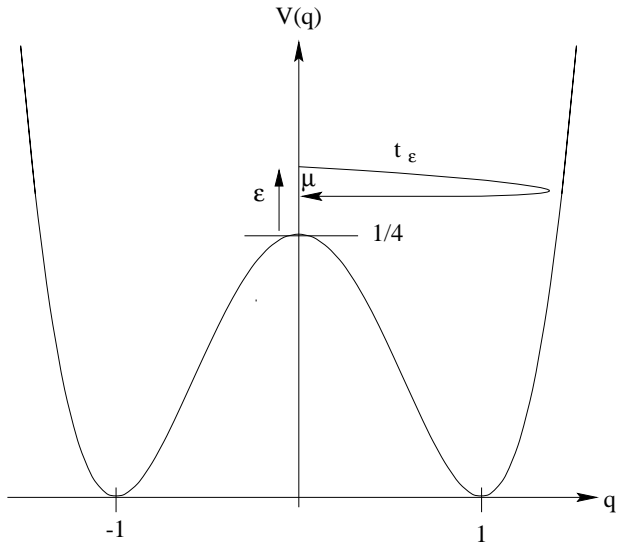}
\end{center}
\caption
{
Potential energy for the Kramers problem as given in Eq.~(\ref{potential})
with $V_0=1$.  The minima of the potential occur at $q=\pm 1$, the maximum
at $q=0$, and the barrier height is $1/4$.  $\varepsilon$ measures the
energy above the barrier, $t_\varepsilon$ is the time it takes a particle
of energy $\varepsilon$ above the barrier starting at $q=0$ to return to 
$q=0$, and $\mu$ is the net energy loss in this half orbit.  The quantities
$t_\varepsilon$ and $\mu$ are discussed in Section~\ref{energylimorbits}.
}
\label{fig1}
\end{figure}

In this paper we restrict ourselves to the Markovian regime, that is, to
the original Kramers problem, deferring the analysis of the non-Markovian
case to a subsequent publication.  The dynamical equation of interest to
us is thus the generic reaction coordinate problem
\begin{align}
\dot{q}&=\frac{p}{m} 
\notag \\ 
\dot{p}& = -\gamma p - \frac{dV(q)}{dq} +f(t)
\label{generic}
\end{align}
where $q(t)$ is the time-dependent reaction coordinate, a dot denotes a
time derivative, $\gamma$ is the dissipation parameter, $V(q)$ is the
potential energy, and $f(t)$ represents Gaussian
delta-correlated fluctuations that satisfy the fluctuation-dissipation
relation
\begin{equation}
\left< f(t)f(\tau)\right>=2\gamma k_BT\delta(t-\tau).
\end{equation}
$k_B$ is Boltzmann's constant and $T$ is the temperature.
The potential $V(q)$ is a double-well potential that is often (and here as
well) taken to be of the form
\begin{equation}
V(q)=V_0\left(\frac{q^4}{4}-\frac{q^2}{2}+\frac{1}{4}\right) =
\frac{V_0}{4}(q^2-1)^2
\label{potential}
\end{equation}
(see Fig.~\ref{fig1}).  The minima of the potential occur at $q=\pm 1$, the
potential barrier has a maximum at $q=0$, and the barrier height there is
$V_0/4$.  The parameter $V_0$ can be used as the unit of energy, and
henceforth we set it equal to unity.  The barrier height is assumed
to be large compared to the temperature (i.e., $k_BT\ll 1/4$); otherwise
the notion of a barrier crossing process loses its meaning. 

The problem of interest is the rate coefficient $k(t)$
for an ensemble of particles whose positions evolve as
realizations of $q(t)$.  The coefficient $k(t)$ measures the 
mean rate of passage of the ensemble across $q=0$ from one well to
the other.  This crossing rate and, in
particular, its asymptotic value $k\equiv k(\infty)$, is associated
with the rate constant of the process represented by the reaction
coordinate.  One usually focuses on the corrections to the rate obtained
from transition state theory (TST) and therefore writes
\begin{equation}
k(t)=\kappa(t) k^{TST}
\label {k}
\end{equation}
where $k^{TST}$ is the rate obtained from transition state theory for
activated crossing, which for our problem and in our units is
\cite{Hanggi}
\begin{equation}
k^{TST}=\frac{\sqrt{2}}{\pi}e^{-1/4k_BT}.
\end{equation}
The deviations from this rate constant are
then contained in the transmission coefficient $\kappa(t)$.  
The construction of $\kappa(t)$ is discussed in subsequent sections.  
In the remainder of this work we designate the stationary or
asymptotic transmission
coefficient $\kappa(t\rightarrow\infty)$ as $\kappa_{st}$.

\begin{figure}[htb]
\begin{center}
\leavevmode
\epsfxsize = 3.2in
\epsffile{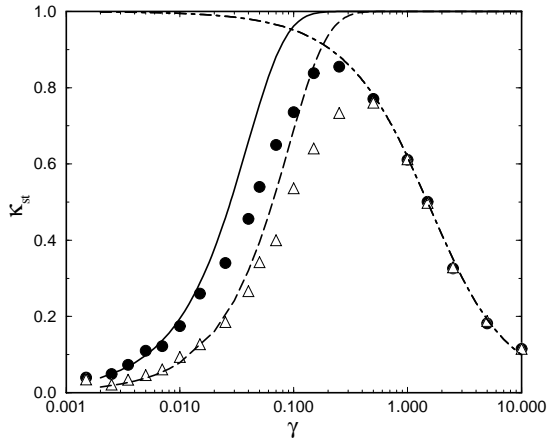}
\end{center}
\caption
{
Transmission coefficient $\kappa_{st}$ vs dissipation parameter $\gamma$
for two temperatures
obtained from direct simulation of Eq.~(\ref{generic}) as described in
Section~\ref{simulations}.  Solid circles: $k_BT = 0.025$; triangles:
$k_BT=0.05$.  As
already predicted by Kramers, the high friction result (short-dashed
curve) is
independent of temperature and equal to $-(\gamma/2)
+((\gamma^2/4)+1)^{1/2}$, which approaches
$\gamma^{-1}$ as $\gamma\rightarrow\infty$. 
At very low friction Kramers predicted that $\kappa_{st}$
is proportional to $\gamma/k_BT$.  The dashed and solid curves, based on
our predictions for the behavior of $\kappa_{st}$ beyond the very
earliest linear regime, are discussed in Section~\ref{compare}.  
}
\label{fig2}
\end{figure}

The input parameters of the problem are the dissipation ($\gamma$) and
the temperature ($k_BT$).
The behavior of the stationary-state transmission coefficient
$\kappa_{st}$ as a
function of these parameters, particularly as a function of $\gamma$, was
the subject of a great deal of work in the 80's.  It was already
realized by Kramers that the behavior of $\kappa_{st}$
vs $\gamma$ would exhibit
a maximum (cf. Fig.~\ref{fig2}) since at intermediate and high dissipation
(``diffusion limited regime") the rate constant decreases with increasing
dissipation while at low dissipation (``energy-diffusion limited regime")
it increases with increasing dissipation.  Indeed, Kramers found
that at as $\gamma\rightarrow\infty$, $\kappa_{st}\approx 1/\gamma$ while 
as $\gamma\rightarrow 0$, $\kappa_{st}\sim\gamma/k_BT$.  These
limits were obtained
from separate arguments, however, which therefore gave no
indication of where precisely the ``turnover" (maximum) might occur. 
Subsequent
work attempted to address this problem \cite{Hanggi}.  The other
parameter dependence,
namely, on temperature, has also been the object of study.  For example, it
is understood that the high-dissipation $\kappa_{st}$ is independent of
temperature whereas $\kappa_{st}$ does depend on temperature in the
low-dissipation regime.  This is clearly seen in the simulation results
in Fig.~\ref{fig2}.  Discussion of Fig.~\ref{fig2} will be resumed later.

The analysis of KT has carried the problem further in that
they actually dealt with the full time-dependent evolution of the
transmission coefficient.  They derived an expression for $\kappa(t)$ in
terms of the time-dependent phase space density, assumed a particular form
for this density, and studied how $\kappa(t)$ goes to its steady state value
as the phase space density relaxes to equilibrium.  In carrying out this
program, however, they relied on approximations that are appropriate for
the diffusion limited regime and therefore were able to calculate
$\kappa(t)$ only in this regime.  They did not check their time-dependent
results against simulations, although they did confirm numerically that
the (temperature-independent) steady state values for the transmission
coefficient that they predict in this regime are satisfactory.  They also
did not deal with the energy-diffusion limited regime.  

Thus, to complement the work of KT in the Markovian regime
we accomplish three goals in this paper:

\begin{enumerate}
\item
We carry out simulations of the time-dependent transmission
coefficient to assess
and confirm the validity of the KT formulation in the diffusion-limited
regime.  Their results agree very well with simulations in this
regime. 

\item
We formulate the complementary theory for the time-dependent
transmission coefficient
in the energy-diffusion limited regime and compare our results with
numerical simulations in this regime.  Our theory captures the complex
oscillatory and stationary state behavior of $\kappa(t)$ in this
regime very well.

\item
We calculate and check via our simulation results the temperature
dependence of the asymptotic transmission coefficient in the energy-limited
diffusion regime beyond the Kramers result.

\end{enumerate}

In Section~\ref{sec2} we briefly review the reactive flux formalism.  
Section~\ref{simulations} contains a short description of the numerical
methods used in our simulations. 
The results of KT in the diffusion-limited regime are recalled and compared
to our simulation results in Section~\ref{difflim}.  The departure of the
KT results from the simulations at low dissipation are illustrated in this
section.  In Sections~\ref{energylimorbits} and \ref{energylimtranscoeff}
we present our theory for the time-dependent transmission coefficient
$\kappa(t)$ and its asymptotic value in the energy-diffusion limited regime.
The comparison of our results with those of numerical simulations is
presented in Section~\ref{compare}.  We conclude in
Section~\ref{conclusion}.

\section{Reactive Flux Formalism}
\label{sec2}
Two decades ago saw the development of the reactive flux formalism for
the rate constant over a barrier \cite{Hanggi,Strauba,Straubb}. 
This formalism was
important because it
made possible efficient numerical simulations of the rate constant without
having to wait the inordinately long time that it takes a particle at
the bottom of one well to climb up to the top of the barrier.  In the
reactive flux method the problem is formulated in terms of the
particles of the thermal distribution that are sufficiently energetic
to be above the barrier at
the outset.  In this way, it is not necessary to wait for
low-energy particles (the vast majority of all the particles)
to first acquire sufficiently high energies via thermal fluctuations.  

We follow the notation of KT. The
rate constant $k(t)$ in the reactive flux formalism is
\begin{align}
k(t)&= \frac{\left< \dot{\theta}_P (q_0)\theta_P[q(q_0,
v_0,t]\right>}{\left< \theta_R(q_0)\right>}
\notag
\\ \notag \\
&=\frac{\left< v_0\delta(q_0)\theta_P[q(q_0,v_0,t)]\right>}{\left<
\theta_R(q_0)\right>},
\label{flux}
\end{align}
where the top of the barrier is at position $q_0=0$, $\theta_R(x)=1$
if $x < 0$ and $0$ otherwise, and $\theta_P=1-\theta_R$.  The brackets
$\left<~\right>$ represents an average over initial equilibrium conditions.
With the particular choices made in Eq.~(\ref{flux}) one is calculating
the transition rate {\em from} the left well {\em to} the right.
KT proceed through a series of steps that finally yield 
for the transmission coefficient introduced in Eq.~(\ref{k}) the relation
\begin{equation}
\kappa(t) = \frac{m}{k_B T}\int_{-\infty}^\infty dv_0 v_0
e^{-mv_0^2/2k_BT}\chi(v_0,t)
\label{transmission}
\end{equation}
where
\begin{equation}
\chi(v_0,t) = \int_0^\infty dq \int_{-\infty}^\infty dv W(q,v,t; q_0=0,v_0)
\label{chiv}
\end{equation}
and $W(q,v,t;q_0=0,v_0)$ is the conditional phase space distribution
function that corresponds to an ensemble of particles starting at
$(q_0=0,v_0)$ at $t=0$.

The (nonequilibrium) conditional probability distribution $W$
clearly lies at the crux of the calculation: it is the distribution
associated with a Langevin equation or a generalized Langevin equation that
describes the evolution of the position $q(t)$ and momentum $p(t)=mv(t)$
of the ensemble of particles.  In the problem of interest here, $W$ is the
distribution associated with Eq.~(\ref{generic}), that is, it is the
solution of the Fokker-Planck equation (we set the mass equal to unity)
\begin{equation}
\frac{\partial W}{\partial t} = 
-\frac{\partial}{\partial q}(pW) +
\frac{\partial}{\partial p} \left[ (\gamma p + \frac{dV(q)}{dq})W\right] +
\gamma k_B T \frac{\partial^2 W}{\partial p^2}.
\label{fpqp}
\end{equation}
with appropriate initial and boundary conditions.

\section{Simulations}
\label{simulations}

The numerical solution of Eqs.~(\ref{generic}) is performed according to
the following main steps. The integration is carried out using the
second order Heun's
algorithm \cite{Gard}, which has been tested in
different stochastic problems with very reliable results \cite{Toral94}.
A very small time step is used, ranging from
$0.001$ to $0.0001$, as in
Ref.~\cite{Strauba}.  The numerical evaluation of the
transmission coefficient $\kappa(t)$ follows the
description of Refs.~\cite{Strauba,Straubb}. We start the simulation with
$N$ particles 
(either 4000 or 1000 depending of the circumstances), all of them above
the barrier at $q=0$, half with a positive velocity
distributed according to the Boltzmann distribution in energy, which
translates to the velocity distribution $v\exp(-v^2/2 k_B T)$,
and the other half with the same distribution but with negative velocities.

The transmission coefficient is extracted from these sets of simulated
data by calculating \cite{Strauba}
\begin{equation}
\kappa(t) = \frac{N_+ (t)}{N_+(0)} - \frac{N_- (t)}{N_-(0)},
\label{extract}
\end{equation}
where $N_+(t)$ and $N_-(t)$ are the particles that started with positive
velocities and negative velocities respectively and
at time $t$ are in or over the right hand well (i.e. the particles for which
$q(t)>0$).

\begin{figure}[htb]
\begin{center}
\leavevmode
\epsfxsize = 3.6in
\epsffile{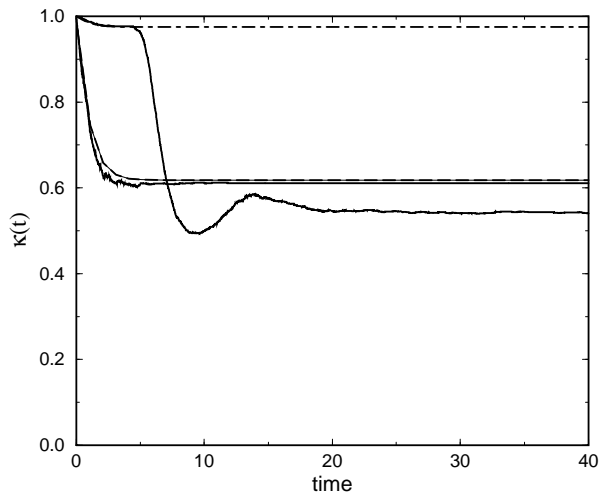}
\end{center}
\caption
{
Transmission coefficient $\kappa(t)$ vs time for two values of the
dissipation parameter $\gamma$ and for temperature $k_BT=0.025$.
The solid curves are the results of our numerical simulations.  
For high dissipation ($\gamma=1.0$, on the right of the maximum in
Fig.~\ref{fig2}),
the transmission coefficient decreases monotonically.  For low dissipation
($\gamma=0.05$, on the left of the maximum in Fig.~\ref{fig2})
the decay is oscillatory. 
The dashed curves are those calculated from the
theory of KT (cf. Eq.~(\ref{KTdifflim}), which is seen to predict
the numerical
results very well for high dissipation, the regime for which
the theory is designed.  The agreement seen in the figure
is typical for large dissipation
parameters. Not surprisingly, the theory fails at low dissipation. 
}
\label{fig3}
\end{figure}

Figure~\ref{fig3} exhibits examples of the
temporal behavior of the transmission coefficient, Eq.~(\ref{extract}),
for two typical different parameter regimes as described
in the figure caption.  Of interest in this paper is the low-dissipation
oscillatory behavior, which is illustrated and discussed
further in subsequent sections.

As discussed by Straub et al.
\cite{Straubb}, whereas the exact transmission coefficient
reaches a constant non-zero value that corresponds to the equilibrium
transition rate for the problem, the transmission coefficient calculated
using the reactive flux method flattens out but continues to decrease
with time.  If the temperature is low, this decrease is
slow and the results appear flat on the sort of time scale shown in
Fig.~\ref{fig2}.  
In this case the value of $\kappa(t)$ in this flat region is identified
with $\kappa_{st}$.  If the temperature is not so low, then the
decaying tail
is extrapolated back to its intersection with the vertical
axis according to the relation
\begin{equation}
\kappa(t) \sim \kappa_{st} e^{- Kt}
\label{discrepancy}
\end{equation}
where $K$ is the decay constant of the tail.  
Thus, the value of the intersection is identified as $\kappa_{st}$.

The numerical values of $\kappa_{st}$ presented in Fig.~\ref{fig2}
are obtained according to these schemes.  Numerical uncertainties
are fairly large and difficult to avoid for small values of $\kappa_{st}$.

\section{Diffusion Limited Regime}
\label{difflim}

Equation~(\ref{fpqp}) can not be solved exactly for the bistable potential,
and therefore the distribution $W$ is the quantity in the calculation of
the transmission coefficient that requires approximation.
The quality of the result of course depends on the quality of the
approximations. KT focus on the regime of moderate to high
dissipation, that is, on the Grote-Hynes model.
In this regime it is appropriate to adapt the theory of Adelman
\cite{Adelman} to the assumption that the barrier is parabolic.  
With this approximation for the conditional probability, KT
derive expressions for $\kappa(t)$ from which the Grote-Hynes
results are recovered in the long-time limit.  Their result is
\begin{equation}
\kappa(t)=\frac{e^{\mu_1t}-e^{\mu_2t}}{\left(
e^{2\mu_1t}\frac{\omega^2}{\mu_1^2} +e^{2\mu_2t}\frac{\mu_1^2}{\omega^2}
+2e^{-\gamma t}-4\frac{\omega_1^2}{\omega^2}\right)^{1/2}}
\label{KTdifflim}
\end{equation} 
where
\begin{equation}
\omega_1\equiv \sqrt{\frac{\gamma^2}{4}+\omega^2},\qquad
\mu_{1,2}=-\frac{\gamma}{2} \pm \omega_1,
\end{equation}
and $\omega$ is the ``frequency" of the top of the barrier, which for the
potential (\ref{potential}) is unity.

In order to assess the KT results and their regime of validity, we
point again to Fig.~\ref{fig2}, where we show the asymptotic
transmission coefficient as a function of
the dissipation parameter obtained from simulations based on
Eq.~(\ref{transmission}).  In this
figure we have exhibited the results for two temperatures. 
The high dissipation side of the curve is independent of
temperature, but the low dissipation transmission coefficient is inversely
proportional to temperature.  The short-dashed curve is the
Kramers/Grote-Hynes result
(and therefore also the result recovered by KT).  Clearly, it is excellent
at high dissipation (a well-known result), but does
not capture the turnover and subsequent decrease in the
transmission coefficient as one enters the energy-diffusion
limited regime.

Figure~\ref{fig3} exhibits a typical simulation result in the high
dissipation regime ($\gamma=1.0$) (monotonic somewhat jagged curve) and
also the associated KT prediction Eq.~(\ref{KTdifflim}) (dashed curve). 
The agreement is seen to be very good over the
full time-dependence of
the transmission coefficient. The very small discrepancy in the
asymptotic regime is also typical, with the theoretical results slightly
above those of the simulation.  This
may in part be due to the issues surrounding
Eq.~(\ref{discrepancy}).  In any case, this degree of agreement
between the
KT theory and the simulations is typical in the regime well
beyond the turnover in Fig.~\ref{fig2}. 
The theory fails, however, at low dissipation
parameters.  In Fig.~\ref{fig3} we show typical results
in this regime -- these are for $\gamma=0.05$ and $k_BT=0.025$. 
Although the two values of $\gamma$ that we have chosen for this figure
lead to similar values of the stationary transmission coefficient, the
decay towards these values is clearly very different in the two cases.
The simulation results (continuous curve) and the KT result
(dotted curve) agree at early times, until the simulation results
rather suddenly
drop and settle in an oscillatory way to a completely different lower
asymptotic value than the theoretical result \cite{Montg}. 
The KT result in fact 
asymptotes to the Grote-Hynes rate, which in this regime is not correct.
In the next sections we obtain results that
predict the correct oscillatory and asymptotic behavior of the
transmission coefficient in the very low dissipation regime.

\section{Energy-Diffusion Limited Regime - Orbits}
\label{energylimorbits}

In this section and the next, we detail the approach complementary to
that of KT to
obtain analytic results for the time-dependent transmission coefficient
in the energy-diffusion limited regime.  Our results are compared to
numerical simulations in Section~\ref{compare}.

For our calculation it is convenient to write the transmission
coefficient as (cf. Eq.~(\ref{extract}))
\begin{equation}
\kappa(t) = \kappa_+(t)-\kappa_-(t),
\label{difference}
\end{equation}
where $\kappa_+(t)$ is the fraction of particles that started above the
barrier with positive velocity $v_0>0$ 
whose coordinate $q(t)>0$, i.e., those that are over or in
the right hand well at time $t$ given that they were over the right
hand well
initially, and $\kappa_-(t)$ the fraction that started above the barrier
with negative velocity $v_0<0$ and are over or in the right hand well at
time $t$.

When dissipation is very slow, the particles that start above the barrier
lose their energy very slowly as they orbit around
at an almost constant energy during each orbit 
(recrossing the barrier many times if they start with sufficiently
high energy). One can calculate the approximate energy loss per orbit
and thus keep track of how long it takes a particle of a
given initial energy above the barrier to lose enough energy to be caught
in one well or the other.  There is of course a thermal distribution of
such particles above the barrier, and the time that it takes the ensemble
to lose enough energy to be caught in a well is correspondingly
distributed. 

The principal ingredients in this calculation are thus 
\begin{enumerate}
\item The time that it takes to complete an orbit;
\item The energy loss per orbit;
\item The distribution of times at which particles eventually become
trapped in one well or the other.
\end{enumerate}
In this section we concentrate on the first two of these ingredients.
The third component, which then leads directly to the transmission
coefficient, is considered in the next section.

Because of the slow energy variation it is here convenient to rewrite the
dynamical equations (\ref{generic}) in terms of the displacement and the
{\em energy}
\begin{equation}
E=\frac{p^2}{2} +V(q)
\end{equation}
instead of the momentum \cite{Stratonovich,Ourbook}.
This simple change of
variables leads to
\begin{align}
\dot{q}&=\left\{ 2 \left[E-V(q)\right]\right\}^{1/2} 
\label{genericq}
\\ \notag \\
\dot{E} &= -2\gamma \left[E-V(q)\right] +
f(t)\left\{ 2 \left[E-V(q)\right]\right\}^{1/2}.
\label{genericE}
\end{align}
This set can of course not be solved exactly either, but one can
take advantage of the fact that for small dissipation $\gamma$
the temporal variation of the energy is much slower than that of
the displacement.

The first quantity to calculate (approximately) is the time
that it takes a particle of energy $E$ above the barrier ($E>1/4$) to
complete one passage from $q=0$ to the edge $q_+$ of the potential
well and back to $q=0$ (cf. Fig.~\ref{fig1}).
For this calculation we assume (this is the approximation) that the
energy $E$ remains fixed during this passage.  
$q_+$ is given by the solution of
$E=V(q)$, i.e., by $q_+=(1+\sqrt{4E})^{1/2}$.
We label $t(E)$ this time of travel
between $q=0$ and $q=q_+$ and back to $q=0$ and call this a
``half-orbit." The time
$t(E)$ is then simply obtained
by integrating Eq.~(\ref{genericq}):
\begin{equation}
t(E) = 2\int_0^{q_+} \frac{dq}{\left\{2\left[E-V(q)\right]\right\}^{1/2}}.
\end{equation}
With the potential (\ref{potential}) this can be expressed in terms of
a complete elliptic integral of the first kind, ${\cal{K}}(z)$:
\begin{equation}
t(E) = \frac{2}{(4E)^{1/4}}{\cal{K}}\left(\frac{1+2\sqrt{E}}{4\sqrt{E}}\right).
\label{halforbit}
\end{equation}
For low temperatures the particles that determine the behavior
of the transmission coefficient are primarily those just above 
the top of the barrier.  For these energies an
excellent approximation to Eq.~(\ref{halforbit}) shows that the time
to traverse the potential depends logarithmically on the
excess energy of the particle above the barrier: 
\begin{equation}
t(E) \approx \ln \frac{16}{(E-\frac{1}{4})}.
\label{importanttime1}
\end{equation}
It is convenient for this and subsequent results to introduce the notation
\begin{equation}
\varepsilon\equiv E-1/4,
\end{equation}
that is, the energy {\em above} the barrier.  In
terms of this energy
\begin{equation}
t_\varepsilon \equiv t(E)\approx \ln \frac{16}{\varepsilon}
\label{importanttime}
\end{equation}
(cf. Fig.~\ref{fig1}).
The corrections to Eq.~(\ref{importanttime}) are of
$O(\varepsilon\ln\varepsilon)$.

{\em Equation~(\ref{importanttime}) is one of the essential results of
this section}.

The energy of the particle of course does not in fact
remain constant as the particle orbits
around.  Kramers calculated this energy loss in his theory of
energy-diffusion limited reactions in terms of the action of the particle.
We recover his result using the following physically transparent argument.
To estimate the actual energy loss, say, along a half orbit, in the simplest
approximation one neglects the fluctuations in Eq.~(\ref{genericE})
(since particles at the high end of the energy distribution primarily
lose energy - this overestimates the energy loss) and
integrates over a time interval $t(E)$. 
Note that ignoring the fluctuations is {\em not} tantamount to ``ignoring
the thermal effects," since the principal thermal effects lie in the
initial distribution of energies, which we take into account exactly.
Subsequently we also take account of the fluctuations ignored in this
simplest approximation.  The $q(t)$ dependence in
Eq.~(\ref{genericE}) still poses a problem, though, because it is in
general an incomplete elliptic
integral itself amenable only to numerical integration.  However, since
$q$ changes rapidly compared to $E$, it would appear reasonable to
perform an average of the
$q$-dependent term over a passage from $q=0$ to $q_+$.  This leads to
the approximation
\begin{equation}
\dot{E} = -2\gamma F(E) 
\label{approxE}
\end{equation}
where
\begin{equation}
F(E)\equiv \int_0^{q_+}dq P_E(q)\left[E-V(q)\right]
\label{average}
\end{equation}
and $P_E(q)$ is the probability density that $q(t)=q$. The assumption
that this probability density is in turn proportional
to the inverse of the velocity at $q(t)$, i.e., 
$P_E(q)\propto 1/\dot{q}$ \cite{Stratonovich,Ourbook},
leads with Eq.~(\ref{genericq}) and appropriate normalization to
\begin{equation}
P_E(q)=\frac{\sqrt{2}}{t(E)} \frac{1}{\left[E-V(q)\right]^{1/2}}.
\label{probabilityq}
\end{equation}
The integral in Eq.~(\ref{average}) can then be expressed in terms of
complete elliptic integrals.
For the energies of interest here an excellent approximation is obtained
by retaining the $E$-dependent contribution $t(E)$ in
Eq.~(\ref{probabilityq}) (which is singular as $E\rightarrow 1/4$) but to
set $E=1/4$, the barrier height, elsewhere in the integrand.  We thus
obtain
\begin{equation}
F(E)\approx \frac{2}{3}\ln \frac{16}{(E-\frac{1}{4})} \approx
\frac{2}{3t(E)},
\label{approxF}
\end{equation}
Thus, for a first estimate of the energy loss per half orbit one might use
Eq.~(\ref{approxE}) with (\ref{approxF}): 
\begin{equation}
\dot{E} = -\frac{4\gamma}{3}\frac {1}{t(E)} 
\label{approxapproxE}
\end{equation}

The next question then is: on the basis of the approximate equation
(\ref{approxapproxE}), how much energy does a particle that begins
with energy
$E$ actually lose per half orbit?  The answer of course depends on the
energy -- in particular, according to Eq.~(\ref{approxapproxE}),
it depends on the time that it takes that particle to complete its
half orbit. Let $\mu(E)$
be the energy loss in a half orbit of a particle that has initial energy
$E$.  The deterministic equation Eq.~(\ref{approxapproxE})
can be integrated directly 
and to give in terms of the energy $\varepsilon$
above the barrier
\begin{equation}
\left( \frac{4\gamma}{3}-\mu\right)\ln\frac{16}{\varepsilon} = 
\left( \varepsilon-\mu\right)\ln \frac{\varepsilon-\mu}
{\varepsilon} +\mu.
\label{intermediate}
\end{equation}
The solution $\mu(\varepsilon)$ of this equation clearly
depends on the initial energy $\varepsilon$ of the particle above the
barrier.
It is in fact not difficult to see that $\mu(\varepsilon)$ increases
with $\varepsilon$.
However, this dependence is sufficiently mild that 
the error made in ignoring this
variation is no larger than the effect of the corrections of
$O(\varepsilon\ln\varepsilon)$ that have been ignored in
Eq.~(\ref{importanttime}).
We can therefore pick a convenient energy above the barrier to calculate
this loss in a half orbit.  An upper bound to the solution is obtained
by setting $\varepsilon >> \mu$ in Eq.~(\ref{intermediate}),
which immediately yields the simple relation
between the energy loss per half orbit and the dissipation parameter
\begin{equation}
\mu=\frac{4}{3}\gamma.
\label{importantenergy}
\end{equation}
This result agrees with that obtained via weak-collision
arguments or small dissipation arguments by a variety of essentially
equivalent routes.  At the other extreme,
we can set $\varepsilon=\mu$.  Particles with this energy
have just enough energy to complete
a half orbit from $q=0$ to $q_+$ and back to the origin. Solution of
Eq.~(\ref{intermediate}) in this case leads to a $\mu$
that is somewhat smaller than Eq.~(\ref{importantenergy}),
by approximately 10-15\% for dissipation parameters in the energy-diffusion
limited regime. 

We shall call the value Eq.~(\ref{importanttime}) of $\mu$ the
``bare" energy loss per half orbit and keep in mind that it is an
upper bound for $\mu$.

{\em Equation~(\ref{importantenergy}) is the second
important result of this section.}

\begin{figure}[htb]
\begin{center}
\leavevmode
\epsfxsize = 3.8in
\epsffile{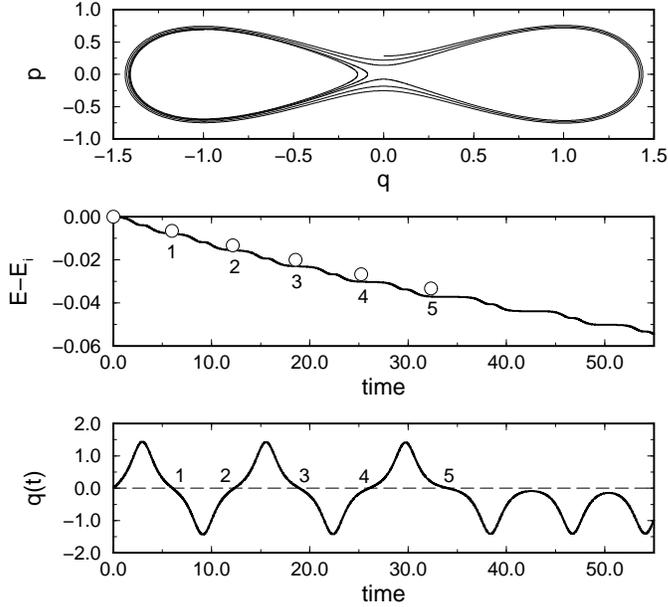}
\end{center}
\caption
{
Simulation of the evolution of a single particle according to 
Eqs.~(\ref{genericq}) and (\ref{genericE}) with no fluctuations.  The
particle starts at $q=0$ moving toward the right and with
initial energy $E_i = 0.254$ ($\varepsilon_i=0.04$).  The dissipation
parameter is $\gamma=0.005$.  Top panel: phase space portrait.  Middle
panel: the decay of the energy (solid curve).  Bottom panel: the particle
trajectory.  The times at which the particle crosses $q=0$ have been
labeled consecutively -- in this case there are five such times before the
particle becomes trapped. Circular symbols: the decay of the energy
according to our approximation. The symbols are placed at the theoretical
barrier recrossing times.
}
\label{fig4}
\end{figure}

In the absence of fluctuations, the only approximations (to order
$\varepsilon \ln\varepsilon$) have been to average over a half orbit
in calculating the time for the return of a particle to the origin, and
to assume that in the calculation of this period the energy loss of the
particle over the half orbit is negligible.  Thus, we have approximated
the evolution of the energy of a particle as determined by
Eqs.~(\ref{genericq}) and (\ref{genericE}) with $f(t)=0$ by
simply assuming an energy loss of $\mu$ at the end of each half orbit.
In Fig.~\ref{fig4} we show the results of the
exact integration of Eqs.~(\ref{genericq}) and (\ref{genericE}) in the
absence of fluctuations for a particle starting at $q=0$ with an energy
$\varepsilon=0.04$ above the barrier and moving toward the right.
The bottom panel shows the
trajectory of the
particle, $q(t)$.  The particle orbits back and forth
above the barrier several times before getting trapped in the left well.
The top panel shows the associated phase space portrait.  The middle panel
shows the decay of the energy (solid curve).  The decay is not entirely
uniform because the rate of energy loss when the particle has high kinetic
energy (i.e. as it goes over the deepest portions of the wells) is
greater than when its kinetic energy is low.  However, this non-uniformity
is very mild.  Our approximation is given by the circular symbols:
we assume an energy loss of $\mu$ per half
orbit, and the symbols are placed at the times of completion of each half
orbit as predicted by our theory. Once the particle gets trapped,
its orbits become smaller and the energy loss per orbit smaller, but
these dynamics are in any case not relevant to our problem. 
We conclude that in the absence of fluctuations the assumption of an equal
loss of energy $\mu$ in the course of each half orbit during the recrossing
process is appropriate.  The other theoretical information used
in drawing the symbols in the middle panel of the figure are the times
it takes to complete each of the half orbits. 
The lower panel in Fig.~\ref{fig4} shows
crossings at the times given in the second column of Table~\ref{tbl1}.
Those predicted by Eq.~(\ref{importanttime}) are presented in the fourth
column.  To obtain these numbers we subtract an energy $\mu=4\gamma/3$
after each half orbit and calculate the appropriate $t_\varepsilon$ for
the next half orbit. The agreement is clearly quite good.

\begin{figure}[htb]
\begin{center}
\leavevmode
\epsfxsize = 3.8in
\epsffile{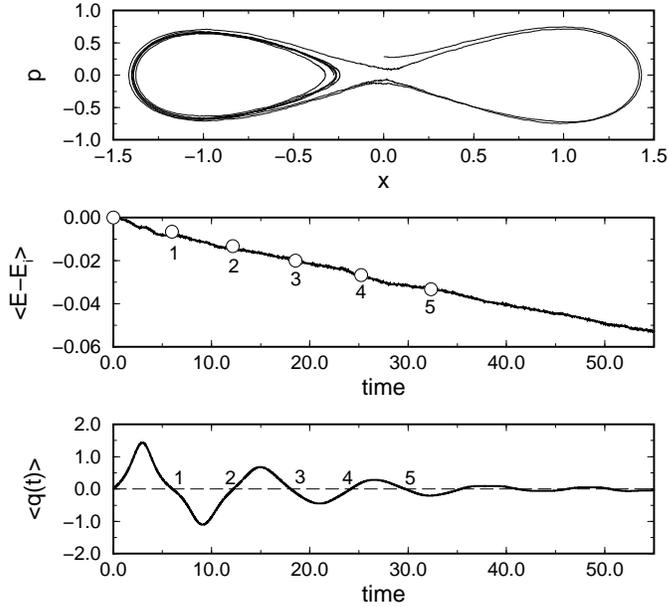}
\end{center}
\caption
{
Simulation of the evolution of $500$ particles according to 
Eqs.~(\ref{genericq}) and (\ref{genericE}) with fluctuations corresponding
to a temperature $k_BT=0.025$.  The
initial energy of each particle above the barrier is $\varepsilon_i=0.04$,
and the dissipation
parameter is $\gamma=0.005$.  Top panel: phase space portrait of a typical
particle.  Middle
panel: the decay of the average energy (solid curve).  Bottom panel:
the average particle trajectory. 
The times at which the particles on average cross $q=0$ have been
labeled consecutively -- in this case there are five such times before the
particle becomes trapped. Circular symbols: the decay of the energy
Circular symbols: the decay of the energy
according to our approximation.  The symbols are placed at the theoretical
barrier recrossing times.
}
\label{fig5}
\end{figure}

Consider now the evolution of particles in the presence of thermal
effects.  In Fig.~\ref{fig5} we again show the results of the
exact integration of Eqs.~(\ref{genericq}) and (\ref{genericE}), but
now in the presence of fluctuations corresponding to a temperature 
$k_BT=0.025$. The second and third panels represent ensemble averages
over $500$ particles all
starting at $q=0$ and moving toward the right with an initial energy
$\varepsilon=0.04$ above the
barrier.  The ensemble average is over different realizations of the
fluctuations.  The bottom panel shows the average trajectory of the
particles, $\left<q(t)\right>$.  On average, the particles orbit back
and forth above the barrier several times.  About half of the
particles eventually get
trapped in the left well and the other half in the right, whence the 
average trajectory settles down to values near zero.
The top panel shows the associated phase space portrait for a typical
particle of the ensemble.  The middle panel
shows the decay of the average energy (solid curve).  The decay is
a bit noisier than that of the deterministic case of Fig.~\ref{fig4}
because the stochastic ensemble is not very large (the fluctuations are of
order $O(N^{-1/2})$ where $N$ is the number of particles).
Our approximation (which neglects thermal effects) is again given
by the circular symbols:
we have again assumed an energy loss of $\mu$ per half
orbit, and the symbols are placed at the times of completion of each half
orbit as predicted by our theory.  Again, the agreement is excellent.
We conclude that on these time scales even in the presence of
fluctuations the assumption of an equal
loss of energy $\mu$ in the course of each half orbit during the recrossing
process is appropriate, and that this value of $\mu$ is essentially
that obtained disregarding the fluctuations. 
Again, the other theoretical information used
in drawing the symbols in the figure are the times it takes
to complete each of the half orbits. 
The crossings observed in the lower panel in Fig.~\ref{fig5} occur
at the times shown in the third column of Table~\ref{tbl1}.
Note that the crossing times are now just a little shorter than in the
deterministic simulation, an indication that the particles have a bit more
energy as they orbit and are thus moving a bit faster. This in turn
indicates that the energy loss per orbit here is actually a bit smaller
than in the deterministic case (just barely perceptibly so on the energy
scale of the middle panels in the figures).  We interpret this as the
effect of thermal fluctuations that counteract the energy loss.  On these
short time scales the thermal fluctuations barely begin to have an effect
on the evolving particles, and hence their behavior is very well captured
by the dissipative component. 
We will see subsequently
that although the short time behavior (such as shown in these figures)
is well captured by an energy loss rate that is independent of the
fluctuations, at longer times the fluctuations play a more important
role and the corresponding $\mu$ should be appropriately renormalized.

To estimate the correction to Eq.~(\ref{importantenergy}) due to the thermal
fluctuations we return to the full set Eqs.~(\ref{genericq}) and 
(\ref{genericE}) and consider the Fokker-Planck equation associated with this 
set for the probability density $W(q,E,t|q_0=0,E_0)$:
\begin{align}
\frac{\partial}{\partial t} W=&-\frac{\partial}{\partial q} \left\{ 
2[E-V(q)]\right\}^{1/2}W +2\gamma \frac{\partial}{\partial E} \left[ E-V(q)
\right]W
\notag\\
\notag\\
&+\gamma k_BT \frac{\partial}{\partial E}\left\{2[E-V(q)]\right\}^{1/2}
\frac{\partial}{\partial E}\left\{2[E-V(q)]\right\}^{1/2}W.
\label{fullFP}
\end{align}
To find an equation for the evolution of the average energy
$\left<E(t)\right>$ we multiply Eq.~(\ref{fullFP}) by $E$ on the left and
integrate over $q$ and $E$.  Upon integration by parts all boundary terms
vanish and one is left with
\begin{equation}
\frac{d}{dt} \left<E\right> = -2\gamma \left< E-V(q)\right> +\gamma k_B T
\label{almost}
\end{equation}
where
\begin{equation}
\left< f(q,E)\right> \equiv \int_0^\infty dE \int dq W(q,E,t|q_0=0,E_0)f(q,E)
\label{defineaverage}
\end{equation}
and the integral over $q$ is over the region where $V(q)<E$.  A well-known
generalization of the argument surrounding Eq.~(\ref{probabilityq}) in the
energy-diffusion limited regime is to assume that the probability density
can be separated in the form \cite{Stratonovich,Ourbook}
\begin{equation}
W(q,E,t|q_0=0,E_0)\approx w(E,t|E_0)P_E(q) 
\end{equation}
(the steps to obtain Eq.~(\ref{intermediate}) are consistent with this
assumption in the special case that $w$ evolves deterministically).  The
average of Eq.~(\ref{almost}) with respect to $q$ is then equivalent to
the earlier average over a half-orbit at fixed $E$. With the assumption
that we can replace $\left<t(E)^{-1}\right>$ with $t(\left<E\right>)^{-1}$
this finally leaves us
with the evolution equation for the mean energy that modifies 
(\ref{approxapproxE}) in that it takes thermal fluctuations into account:  
\begin{equation}
\dot{\left<E\right>} = -\frac{4\gamma}{3}\frac {1}{t(\left<E\right>)}
+\gamma k_BT.
\label{withtemp}
\end{equation}

\begin{table}
\begin{center}
\begin{tabular}{ | l | r | r | r |}
\hline
&Simulation&Simulation&Theory\\ \hline
&Deterministic&$k_BT=0.025$& \\ \hline
$t_0$ & 0.00 & 0.00 & 0.00 \\ \hline
$t_1$ & 6.03 & 6.04& 5.99 \\ \hline
$t_2$ & 12.32 & 12.20& 12.16 \\ \hline
$t_3$ & 18.94 & 18.06& 18.56 \\ \hline
$t_4$ & 26.03 & 24.15& 25.24 \\ \hline
$t_5$ & 34.04 & 29.70& 32.33 \\ \hline
\end{tabular}
\caption{\protect{Zero crossing times of the orbits shown in
Figs.~\ref{fig4} and \ref{fig5}.  Second column: simulation results of
Fig.~\ref{fig4}.  Third column: simulation results of Fig.~\ref{fig5}.
Fourth column: theoretical results.}}
\label{tbl1}
\end{center}
\end{table}

The solution of this equation again clearly
depends on the initial energy of the particle, and the arguments 
surrounding the choice (\ref{importantenergy}) still apply.
An exact explicit solution of Eq.~(\ref{withtemp}) (even to the orders
of approximation considered here) is not possible, but a rough
approximation for low temperatures is
\begin{equation}
\mu(T)= \frac{4}{3}\gamma -\gamma k_BT \ln \frac{12}{\gamma}.
\label{mostimportantenergy}
\end{equation}
Of course, numerical integration of Eq.~(\ref{withtemp}) is trivial, and
in our further analysis we use the outcome of such integration.
For example, for $\gamma=0.005$, a value used in a number of our
simulations, Eq.~(\ref{withtemp}) leads to $\mu(T)=0.004907$
for $k_BT=0.025$ and to $\mu(T) = 0.004485$ for $k_BT=0.035$.  These
are to be compared with the
estimate in Eq.~(\ref{importantenergy}), which is $\mu=0.006667$.
Although at short times we have seen that the ``bare" value of $\mu$ yields
excellent agreement with simulations, these ``renormalized" values do play
a more important role at long times, as will be seen in the next section.

{\em The expression (\ref{mostimportantenergy}) or, more precisely, the
result of integrating Eq.~(\ref{withtemp}), is the third essential
result of this section}.

\section{Energy-Diffusion Limited Regime - Transmission Coefficient}
\label{energylimtranscoeff}
The calculation of the time-dependent transmission coefficient, the
goal of this work, is based on the principal results of
the previous section, namely, the time $t(E)\equiv t_\varepsilon$
it takes a particle
of energy $\varepsilon$ above the barrier to complete a half orbit,
and the energy loss $\mu$ in the course of this
time. In order to carry out this calculation, we must keep track of
where the particles that start above the barrier are at any time
before they get trapped in one or the other of the wells,  and at what time
each particle becomes trapped in one well.  These quantities of course
depend on the initial energy of the particle, which is in turn distributed
thermally.  

The accounting that leads to the result for the transmission coefficient
$\kappa(t)$ turns out to be more transparent if we note that due to
the symmetry of the problem the following interpretation of the terms in
Eq.~(\ref{difference}) leads precisely to the same result as the
description presented there.  We can
imagine that {\em all} particles are forced to start with a positive
velocity $v_0>0$.  Then $\kappa_+(t)$ is, as before, the fraction of
particles that at time $t$ are over or in the right hand well, but now
$\kappa_-(t)$ can be interpreted as the fraction of particles that at time
$t$ are over or in the left hand well.  Furthermore, with this
interpretation it is clear that
\begin{equation}
\kappa_-(t)=1-\kappa_+(t)
\label{interpretation}
\end{equation}
so it is sufficient to follow one or the other.

With this interpretation we thus imagine that all particles above the
barrier start at $q_0=0$ and with a positive velocity $v_0$ distributed
according to the Boltzmann distribution.   Since the energy loss per half
orbit is $\mu$ essentially independently of the energy of the particles
(subject to the approximations and conditions discussed in the last
section), keeping track of particles is conveniently done in ``layers" of
energy thickness $\mu$.  Although the energy loss per half orbit is
essentially independent of energy, the time for a half orbit is not, since
particles of higher energy return to $q=0$ faster than
those of lower energy.  This time dispersion must be accounted for carefully
in constructing $\kappa(t)$.  

Initially the particles are distributed according to a Boltzmann
distribution at temperature $T$.  The fraction of particles
that are above the barrier is 
\begin{equation} 
P(E>1/4) = \frac{1}{k_BT}\int_{1/4}^\infty e^{-E/k_BT} = e^{-1/4k_BT}.
\label{initially}
\end{equation}
We are interested in keeping track of only this fraction and, in
particular, of the fraction of this fraction that is over one well or the
other as time proceeds.  Therefore, we normalize this initial fraction to
unity, i.e., the distribution of interest is
\begin{equation}
p(\varepsilon) = \frac{1}{k_BT}e^{-\varepsilon/k_BT},
\quad 0<\varepsilon<\infty.
\end{equation}

Initially all particles are over the right hand well, i.e.,
\begin{equation}
\kappa_+(0) = \int_0^\infty d\varepsilon p(\varepsilon) = 1,
\qquad \kappa_-(0)=0
\end{equation}
and therefore
\begin{equation}
\kappa(0)=1.
\end{equation}
All those particles that start with an energy between $0$ and $\mu$ above
the barrier never leave the right hand well
since by the time they return to $q=0$ they will have lost energy $\mu$ and
will thus be below the top of the barrier and unable to move out of the
well.  The remainder, those of energy greater than $\mu$ above the barrier,
$\varepsilon > \mu$, do come back and recross the barrier to the
other side.  However, they do so at different times depending on their
energy.  We can write this as follows:  
\begin{align}
\kappa_+(t) &= 1-\frac{1}{k_BT}\int_\mu^{\infty} d\varepsilon
e^{-\varepsilon/k_BT}\theta(t-t_\varepsilon) + \cdots,
\label{round1} 
\end{align}
where the Heaviside theta function $\theta(x) =1$ for $x>0$ and $=0$
for $x<0$.
The theta function insures that recrossing occurs at the right time
for that energy and no sooner.  The $\cdots$ indicate that the calculation
is not yet complete, since we need to continue to keep track of particles
that recross back from the left well to the right, and so on. 
The theta function in Eq.~(\ref{round1})
can be implemented by adjusting the limits of integration: 
\begin{alignat}{2}
\kappa_+(t) &= 1-\frac{1}{k_BT}\int_{f_1(t)}^{16} d\varepsilon
e^{-\varepsilon/k_BT} + \cdots, && \qquad 0<t<\ln\frac{16}{\mu},
\notag\\ \notag\\
&= 1-\frac{1}{k_BT}\int_{\mu}^{16} d\varepsilon
e^{-\varepsilon/k_BT} + \cdots, &&\qquad \ln \frac{16}{\mu}<t<\cdots.
\label{round2}
\end{alignat}
Here $f_1(t)$ is the inverse of the half orbit time, that is,
$\varepsilon=f_1(t)$ is the solution of the relation 
\begin{equation}
t=t_\varepsilon = \ln\frac{16}{\varepsilon},
\label{one} 
\end{equation}
which immediately leads to
\begin{equation}
f_1(t) = 16e^{-t}.
\label{f1}
\end{equation}
Note that the upper limit of the range of the time variable in the first
line of Eq.~(\ref{round2})
is precisely the time it takes a particle of energy $\mu$ above
the barrier to
complete a half orbit, that is, Eq.~(\ref{one}) with $\varepsilon=\mu$. 
During this time all the particles with initial energies
above $\varepsilon=\mu$ will have recrossed the barrier at least once, and
all the particles with energies below $\varepsilon=\mu$ 
are trapped in the right hand well.
The upper limits of integration are written as $16$ (rather
than $\infty$) because 
our approximation for $t_\varepsilon$ (cf. Eqs.~(\ref{halforbit})
and (\ref{importanttime1})) breaks down for higher energies, but
$16$ is, for all practical purposes, infinite when the energy loss
per orbit is of order $\gamma$ and $\gamma$ is very small. In our
subsequent discussion of the time
dependence of the transmission coefficient $\gamma$ is of
order $10^{-2}$.
On the other hand, the time intervals that define 
the ranges of behavior {\em do} involve precisely $\ln(16/\mu)$ because
for small $\mu$ this is the leading contribution to the elliptic integral
discussed in Section~\ref{energylimorbits}.
The integrals in Eq.~(\ref{round2}) can be done explicitly.  Treating
the upper limit as $\infty$ we find
\begin{alignat}{2}
\kappa_+(t)&=
        1-e^{-f_1(t)/k_BT}+\cdots ,
                              & &\qquad
                              0<t<\ln\frac{16}{\mu},
 \notag\\ \notag\\
&=
        1-e^{-\mu/k_BT} +\cdots, && \qquad \ln\frac{16}{\mu}<t<\cdots.
\label{round3}
\end{alignat}
As before, the $\cdots$ indicate that the calculation still continues.

The next stage of the calculation is to follow the particles that recrossed
the barrier once and find themselves over the left well.  Those that
started at $t=0$ with energy in the range $\mu<\varepsilon<2\mu$ above the
barrier and lost energy $\mu$ in their first orbit have entered the region
above the left well with energy between $0$ and $\mu$ above the barrier,
and will therefore never make it out of the left well.  Only those that
started with energy greater than $2\mu$ above the barrier and thus
recrossed the barrier with energy greater than $1/4+\mu$ will be
able to orbit over the left well and return to $q=0$ and eventually
recross once again to the right hand well. 
The following continuation
of Eq.~(\ref{round1}) captures this behavior:
\begin{align}
\kappa_+(t) &= 1-\frac{1}{k_BT}\int_\mu^{\infty} d\varepsilon
e^{-\varepsilon/k_BT}\theta(t-t_\varepsilon) +
\frac{1}{k_BT}\int_{2\mu}^{\infty}
d\varepsilon e^{-\varepsilon/k_BT}\theta(t-t_\varepsilon -
t_{\varepsilon-\mu})
+ \cdots.
\label{round4}
\end{align}
Again, the theta functions insure the correct timing of the recrossings 
and they can be implemented by adjusting the limits of integration:
\begin{alignat}{2}
\kappa_+(t) &= 1-
\frac{1}{k_BT}\int_{f_1(t)}^{16} d\varepsilon
e^{-\varepsilon/k_BT} +
\frac{1}{k_BT}\int_{f_2(t)}^{16} d\varepsilon
e^{-\varepsilon/k_BT}
+ \cdots, & & \qquad 0<t<\ln\frac{16}{\mu},
\notag\\ \notag\\
&= 1- \frac{1}{k_BT}\int_{\mu}^{16} d\varepsilon
e^{-\varepsilon/k_BT} +
\frac{1}{k_BT}\int_{f_2(t)}^{16} d\varepsilon
e^{-\varepsilon/k_BT}
+ \cdots, && \qquad \ln \frac{16}{\mu}<t< 2\ln \frac{16}{(2!)^{1/2}\mu},
\notag\\ \notag\\
&= 1-\frac{1}{k_BT}\int_{\mu}^{16} d\varepsilon e^{-\varepsilon/k_BT}
+\frac{1}{k_BT}\int_{2\mu}^{16} d\varepsilon e^{-\varepsilon/k_BT}
+ \cdots, &&\qquad 2\ln \frac{16}{(2!)^{1/2}\mu}<t<\cdots.
\label{round5}
\end{alignat}
The function $f_2(t)$ is the inverse of the sum of the two half orbit times,
the first half orbit being that over the right hand well with energy
$\varepsilon$, and the second that over the left well with energy
$\varepsilon - \mu$. In other words, $\varepsilon = f_2(t)$ is the solution
of the relation 
\begin{equation}
t=t_\varepsilon + t_{\varepsilon-\mu} =
\ln\frac{16}{\varepsilon} + \ln\frac{16}{\varepsilon-\mu}.
\label{two}
\end{equation}
Exponentiation leads to
\begin{equation}
16^2e^{-t}=\varepsilon(\varepsilon-\mu).
\label{f2}
\end{equation}
This is a quadratic equation that can easily be solved:
\begin{equation}
f_2(t)= \frac{\mu + \sqrt{\mu^2+1024e^{-t}}}{2}.
\end{equation}
The upper limit of the second range of the time variable in
Eq.~(\ref{round5}) is the time it takes a particle of initial energy
$2\mu$ above the barrier to complete two half orbits, that is, 
Eq.~(\ref{two}) with $\varepsilon=2\mu$.  These particles are trapped
in the left well during their second half orbit.  
The integrals can again be carried out explicitly to obtain:
\begin{alignat}{2}
\kappa_+(t)&=
        1-e^{-f_1(t)/k_BT}
	+e^{-f_2(t)/k_BT}
            +\cdots ,
                              & &\qquad
                              0<t<\ln\frac{16}{\mu},
 \notag\\ \notag\\
&=
        1-e^{-\mu/k_BT}+e^{-f_2(t)/k_BT}
      +\cdots, && \qquad \ln\frac{16}{\mu}<t<2\ln
                                       \frac{16}{(2!)^{1/2}\mu},
\notag\\ \notag\\
&=
       1-e^{-\mu/k_BT}+e^{-2\mu/k_BT}
      +\cdots, && \qquad 2 \ln \frac{16}{(2!)^{1/2}\mu}<t<\cdots.
\label{round6}
\end{alignat}

The pattern for $\kappa_+(t)$ and hence for $\kappa(t)$ 
as required in Eq.~(\ref{difference})
has thus been established. We can write the following complete
result for the transmission coefficient as a function of time: 
\begin{alignat}{2}
\kappa(t) &= 1+2\sum_{n=1}^\infty (-1)^n e^{-f_n(t)/k_BT}, && \qquad 
 0<t<\ln\frac{16}{\mu},
\notag\\ \notag\\
&= 1-2e^{-\mu/k_BT} + 2\sum_{n=2}^\infty (-1)^n e^{-f_n(t)/k_BT}, && \qquad
\ln \frac{16}{\mu}<t< 2\ln \frac{16}{(2!)^{1/2}\mu},
\notag\\ \notag\\
&= 1-2e^{-\mu/k_BT} +2e^{-2\mu/k_BT}
+ 2\sum_{n=3}^\infty (-1)^n e^{-f_n(t)/k_BT}, && \qquad
2\ln \frac{16}{(2!)^{1/2}\mu}< t<
3\ln \frac{16}{(3!)^{1/3}\mu},
\label{principal}
\end{alignat}
and so on.
The function $\varepsilon=f_m(t)$ is the solution of the relation
$t=t_\varepsilon + t_{\varepsilon-\mu} +\cdots + t_{\varepsilon-(m-1)\mu}$,
which upon exponentiation turns into the $m^{th}$ order
polynomial equation (a clear generalization of the results Eq.~(\ref{f1})
for $m=1$ and Eq.~(\ref{f2}) for $m=2$):
\begin{equation}
(16)^me^{-t}=\varepsilon[\varepsilon-\mu]\cdots[\varepsilon-(m-1)\mu].
\label{fm}
\end{equation}
The solution of Eq.~(\ref{fm}) can in general not be found in closed form
for $m\geq 3$.  However, an excellent approximation is
\begin{equation}
f_m(t)\approx \left[ m-(m!)^{1/m}\right]\mu +16e^{-t/m}.
\label{fma}
\end{equation}
This form is exact for $m=1$, and it is exactly correct for all $m$ at 
the upper limit of the time range that defines the trapping of the
particles whose initial energy is $\varepsilon=m\mu$.  In other words,
$f_m(t)$ as given in Eq.~(\ref{fma}) is exactly correct at the particular
time $t=t_\mu + t_{2\mu} +\cdots + t_{m\mu}$.

{\em Equation~(\ref{principal}) and its asymptotic limit
\begin{align}
\kappa_{st} & =\kappa(t\rightarrow\infty)=1+2\sum_{n=1}^\infty (-1)^n
e^{-n\mu/k_BT}
\notag\\ \notag\\
&= \tanh \left(\frac{\mu}{2k_BT}\right).
\label{equilibrium}
\end{align}
are the principal results of this paper}. 
These results generalize those of KT in the
Markovian regime to the energy-diffusion limited case. 

It is useful to examine the information contained in the various
contributions to Eq.~(\ref{principal}) and the preceding pieces
that were used to construct it.  The first line contains all first
recrossings of the barrier by all particles that start with sufficient
energy above the barrier to recross is at least once
($\varepsilon > \mu$).
The function $f_1(t)$ accounts for the fact that this first recrossing
occurs at different times for particles of different initial energy,
the last ones (those closest to the barrier that do make it around)
recrossing at time $t=\ln(16/\mu)$.
The particles of higher energy recross first because they are orbiting
at higher velocity, and may be back for
their second recrossing while those of lower energy are still awaiting
their first crossing.  The time distribution of the second recrossing
is contained in $f_2(t)$.  Even while this is occurring, some particles
might already be undergoing their third recrossing, as contained in
$f_3(t)$.  There are fewer and fewer of these faster particles because of
their initial thermal distribution -- this information, too, is contained
in the exponential factors. Of course while all these recrossings are going
on, the particles are losing energy and, depending on their initial energy
and how often they have gone around, they become trapped at sequential
times as expressed in the subsequent lines of Eq.~(\ref{principal}).

We note the nonlinear dependence of $\kappa_{st}$ on the
energy loss parameter $\mu$ and hence on the dissipation parameter
$\gamma$.  The Kramers and other weak collision results in the
energy-diffusion limited regime correspond to the retention of only the
leading term in a small-$\gamma$ expansion of our result,
which then yields $\kappa_{st}\sim\gamma/k_BT$.
 
The detailed behavior of the time-dependent transmission coefficient as
given by Eq.~(\ref{principal}), and its asymptotic value
Eq.~(\ref{equilibrium}), are compared with simulations in the next
section.

\section{Discussion of Results and Comparisons with Numerical Simulations}
\label{compare}

\begin{figure}[htb]
\begin{center}
\leavevmode
\epsfxsize = 4.2in
\epsffile{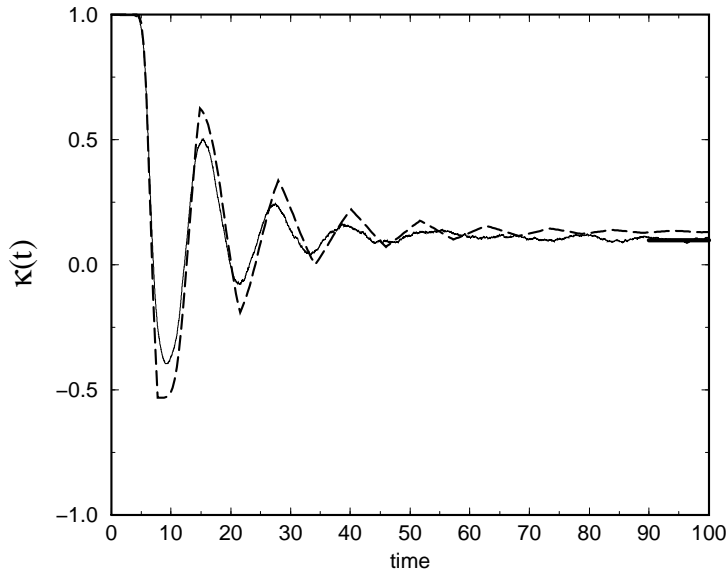}
\end{center}
\caption
{
Transmission coefficient of an ensemble of particles at temperature
$k_BT=0.025$ and dissipation parameter $\gamma=0.005$. 
The particles are initially distributed
above the barrier according to Eq.~(\ref{initially}).
Solid curve: simulation of $4000$ particles.  Dashed curve: our theory,
Eq.~(\ref{principal}), with the bare value $\mu=4\gamma/3=0.006667$.  
Thick line that intersects the right vertical axis: value of the
equilibrium transmission coefficient $\kappa_{st}$ obtained from 
Eq.~(\ref{equilibrium}) using the value of $\mu$ renormalized by thermal
fluctuations, which for this temperature is $\mu(T)=0.004907$. 
}
\label{fig6}
\end{figure}

\begin{figure}[htb]
\begin{center}
\leavevmode
\epsfxsize = 4.2in
\epsffile{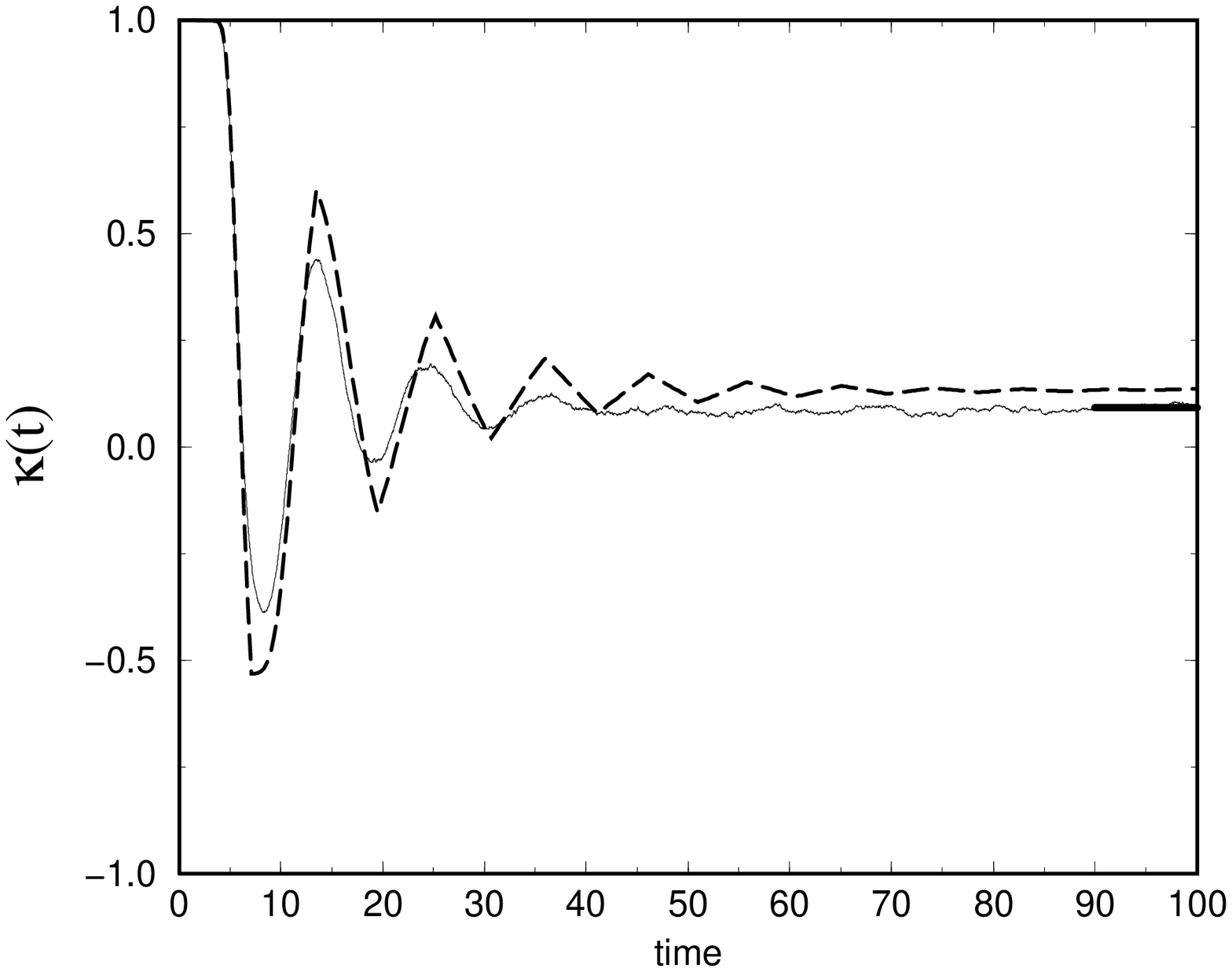}
\end{center}
\caption
{
Transmission coefficient of an ensemble of particles at temperature
$k_BT=0.05$ and dissipation parameter $\gamma=0.01$. 
The particles are initially distributed
above the barrier according to Eq.~(\ref{initially}).
Solid curve: simulation of $4000$ particles.  Dashed curve: our theory,
Eq.~(\ref{principal}), with the bare value $\mu=4\gamma/3=0.01333$.  
Thick line that intersects the right vertical axis: value of the
equilibrium transmission coefficient $\kappa_{st}$ obtained from 
Eq.~(\ref{equilibrium}) using the value of $\mu$ renormalized by thermal
fluctuations, which for this temperature is $\mu(T)=0.007892$. 
}
\label{fig7}
\end{figure}

Figures~\ref{fig6} and \ref{fig7} show the transmission
coefficient $\kappa(t)$ as a
function of time for two sets of parameters.  In Fig.~\ref{fig6}
the dissipation parameter
is $\gamma=0.005$ and the
temperature is $k_BT=0.025$; in Fig.~\ref{fig7} they are 
$\gamma=0.01$ and $k_BT=0.05$.  Note that these are the only
input parameters.  The
solid curves are simulations of an ensemble of 4000 particles.  Each
particle follows the equations of motion (\ref{generic}), and the
transmission coefficient is calculated from the resulting trajectories
according to the reactive flux formalism described earlier.  The
distinctive features of the time dependence are 1) the time at which
the transmission coefficient 
drops rather abruptly from its initial value of unity, and the slope
of this drop; 2) the frequencies
and amplitudes of the oscillations; and 3) the asymptotic value, identified
as the equilibrium value $\kappa_{st}$ of the transmission coefficient.
The dashed curves are the result of our theory, Eq.~(\ref{principal}), with
the bare value $\mu=4\gamma/3$ for the energy loss per half orbit. We stress
that there are no adjustable parameters in these curves.  The agreement
between the theory and simulations is clearly very good in both
cases.  Without
adjustable parameters, the theory captures each of the three 
distinctive features listed above.  Indeed, we stress that with a single
expression we are able to reproduce the temporal behavior of the system
over essentially {\em all} time scales. 

First we discuss those aspects of the theory that work extremely well,
and then we suggest ways in which improvements could be made if even
closer agreement is desired.
The theory essentially {\em exactly} captures the first striking
feature in the transmission coefficient, namely, the time
and slope of the first abrupt drop. The drop
corresponds precisely to the point at which the KT theory
and therefore also the Grote-Hynes theory deviate from
the simulation results in Fig.~(\ref{fig3}) (the parameter values there and
here are different).  The time of the drop and the slope of the first drop
are determined by the particles that first recross from one well to the
other.  Although this time behavior is contained in the full sum of the
first line of Eq.~(\ref{principal}), the first portion of the sum, that is,
the terms shown explicitly in 
Eq.~(\ref{round3}), already capture this behavior very well.  These terms
account for the first crossing over the barrier and do not include the
effects of recrossings (which are contained in the higher $f_m(t)$, as
described earlier).  In other
words, the drop is principally due to the particles near the top of the
barrier that manage to recross. 
The determining features in the early time behavior of $\kappa(t)$ are
thus the time range $0<t<\ln(16/\mu)$,
the value of $\mu$, and the decay function $f_1(t)$. 
Once the least energetic particles are trapped, however, the
particles of initially higher energies (although fewer in number)
sequentially dominate the oscillatory behavior of $\kappa(t)$, that is,
the higher terms in the sum become sequentially important in the time
dependence of $\kappa(t)$ as time increases.  
The subsequent oscillations thus reflect recrossings of the barrier.
These are also captured quite accurately. 
The simulations and the theory settle to asymptotic values that are
extremely close to one another.   

The theory could of course be improved in various ways. First, the
theoretical oscillations are sharper and continue for a longer
time than those of the simulations because the theory averages or
altogether ignores additional effects that lead to particle dephasing.  
For example, the sharp (deterministic) time
cuts (which cause the sharp minima and maxima in the theory) are in
reality somewhat distributed due to the fact that the particles lose energy
continuously and not just at the ends of half orbits as assumed in the
theory, and due to fluctuations that allow energy gains throughout the
dynamics.  Indeed, particles that we take to be trapped forever can in
reality recross the barrier due to thermal fluctuations (but in very
small numbers at low temperatures, as reflected in the small value of
the equilibrium transmission coefficient).
Furthermore, the average energy loss per half orbit,
$\mu(\varepsilon)$, has been assumed independent of $\varepsilon$, but in
reality $\mu$ increases with $\varepsilon$, even
in the ``bare" approximation that ignores thermal fluctuations in the
calculation of $\mu$.  This dependence is contained in
Eq.~(\ref{intermediate}), but we have approximated the solution of
Eq.~(\ref{intermediate}) by the $\varepsilon$-independent
Eq.~(\ref{importantenergy}), which is an upper bound on $\mu$.  
On the other hand, thermal fluctuations
contribute to a decrease in $\mu$ as a function of temperature.  
Indeed, if one analyzes average energy loss trajectories such
as the one shown in the middle panel of Fig.~\ref{fig5} but for slightly
higher initial energies, one finds that there is a small upward
curvature that indicates that $\mu$ starts with the bare value but that
it becomes renormalized by thermal fluctuations as time proceeds. 

The theoretical equilibrium values of the transmission coefficient
achieved by
the dashed curves in Figs.~\ref{fig6} and \ref{fig7} are a little
higher than the simulation
result. Note that the entire dashed curve in both cases has 
been calculated using the upper-bound bare energy loss parameter 
$\mu=4\gamma/3$.
If we rely on our result Eq.~(\ref{equilibrium}), the small
difference between simulated and calculated results would
indicate that the actual value of $\mu$ at long times is a bit lower than
the value assumed in the dashed curve.
We also show in the figures the
equilibrium transmission coefficient calculated in each case according
to Eq.~(\ref{equilibrium}),
but now with the renormalized energy loss parameter that follows from
solving Eq.~(\ref{withtemp}). For the parameters in Fig.~\ref{fig6} this
value is $\mu(T) =0.004907$, and in Fig.~\ref{fig7} it is
$\mu(T)-0.007892$
(thick short lines that intersect the right vertical axis). 
These values
of the stationary transmission coefficient agree extremely with
the simulation results in both cases.

Finally, it is interesting to pursue in more detail the dependences
of the equilibrium
transmission coefficient $\kappa_{st}$ on the input parameters ($\gamma$
and $k_BT$) and, in particular, to evaluate the prediction
Eq.~(\ref{equilibrium}) for these dependences.  As noted earlier, the
existing results correspond to a small-$\gamma$ expansion of our result.
In practically every instance that we have found in the literature, the
equilibrium transmission coefficient vs dissipation parameter is drawn on
a log-log plot that would mask the dependence that we are focusing on.
For this analysis we return to Fig.~\ref{fig2} and, in particular, the low
dissipation portion of the figure.  We present two sets of results for
$\kappa_{st}$ vs $\gamma$ on a semi-log plot, one for
temperature $k_BT=0.025$ (the
temperature used throughout most of this paper) and the other for
$k_BT=0.05$.  The solid circles are the simulation results for the lower
temperature, and the triangles are for the higher temperature.  The first
and very clear point to note is that the $\gamma$ dependence is indeed not
linear beyond the very lowest dissipation parameters.  The
upward curvature with
increasing $\gamma$ is clear.  We then ask whether the $\tanh$ dependence
in Eq.~(\ref{equilibrium}) captures this curvature.  The only
small uncertainty lies in the value of the parameter $\mu$ to use in
answering this question.  For the particular dissipation parameter value
$\gamma=0.005$ and temperature $k_BT=0.025$, and also for $\gamma=0.01$
and $k_BT=0.05$, we have established that
the equilibrium value we
predict is quite accurate if we use the bare $\mu$ and even more
accurate if we use the renormalized value $\mu(T)$.
In Fig.~\ref{fig2} we show the theoretical curve Eq.~(\ref{equilibrium})
for two temperatures
using the renormalized value $\mu(T)$. The solid curve is for
temperature $k_BT=0.025$, and the dashed curve for  
$k_BT=0.05$.  Clearly the theoretical curves capture the
simulation results very well over a substantial range of dissipation
parameters values.

\section{Conclusions}
\label{conclusion}

We have developed a theory for the time dependent transmission coefficient
in the energy-diffusion limited regime of the Kramers problem.  Our work
complements
that of Kohen and Tanner \cite{Kohen}, whose theory covers only the
diffusion limited regime. We
arrive at an explicit analytic prediction for the transmission coefficient
that describes the rate at which particles in the potential of
Fig.~\ref{fig1} move over the barrier from the region associated with
one well to that associated with the other.  Our result depends
{\em only} on
the two input parameters of the problem, the dissipation and the
temperature, and involves no adjustable parameters.
For low dissipation and low temperature, our result
captures the correct behavior on {\em all} time scales,
leading to excellent agreement with simulations in the short time regime
where the transmission coefficient oscillates, and also in the long
time regime
where it settles to an asymptotic equilibrium value.  Our prediction for
the equilibrium value as a function of the dissipation parameter and
temperature extends beyond the regime of existing theories in this regime.

The theory starts with the Langevin equation for the system and proceeds
in two stages. First, we calculate
trajectory features such as the time it takes a particle to complete a
half orbit and the energy loss in the course of a half orbit. 
This provides the
ingredients that allow us to determine when particles cross and recross
the barrier between one well and the other as a function of their initial
energy, and how long it takes for a particle to become trapped in one well
or the other.  The second stage is the combination of these ingredients
for an ensemble of particles that are initially distributed canonically.
The resulting oscillatory early time behavior of the transmission
coefficient reflects barrier recrossing events of initially
energetic particles. The equilibrium value of the transmission coefficient 
reflects the residual steady state flux of particles essentially
trapped in one well or the other.

We also note that we have carried out new simulations in regimes where
theoretical predictions were available but had not been checked, and
in the regimes of interest in this paper where neither theory nor
simulations were previously available.  Through these simulations
it has been possible for us to check our own work as well as that of others
(specifically, the KT results for the diffusion-limited regime), and
we now have available simulation results that can serve as a backdrop
for further theoretical developments.

That a number of chemical reactions occur in the low dissipation regime
has, of course long been clear, not only on the basis of theoretical
considerations but on the basis of a variety of experiments (see e.g.
\cite{Hasha,Troe}). 
The ability to probe reactions on a very short time scale is
much more recent\cite{Zewail,Castleman}.  The theory we have
presented provides a more complete understanding of energy-diffusion
limited reactions than has heretofore been available.  To further
complete the picture, our work continues in
order to generalize our model, most immediately to non-Markovian
processes \cite{Romero}.


\section*{Acknowledgment}
This work was done during a sabbatical leave of J.M.S. at the University
of California, San Diego granted by the Direcci\'o General de Recerca de
la Generalitat de Catalunya (Gaspar de Portol\'a program ).
The work was supported in part by the U.S. Department of Energy under
Grant No. DE-FG03-86ER13606, and in part by the Comisi\'on
Interministerial de Ciencia y Tecnolog\'ia (Spain)
Project No. DGICYT PB96-0241.

\end{spacing}

\end{document}